\documentstyle[pra,aps,twocolumn,psfig,epsf]{revtex}

\newcommand{\bra}[1]{\mbox{$\langle #1 |$}}
\newcommand{\ket}[1]{\mbox{$| #1 \rangle$}}
\newcommand{\bracket}[2]{\mbox{$\langle {{#1}} \mathrel{ | {\vphantom
        {{#1} {#2}}} \kern-\nulldelimiterspace} {{#2}} \rangle$}}

\newcommand{\rem}[1]{}

\newcommand{\ros}{\rho_s}

\newcommand{\rmes}{\rho_s^{(+)}}

\newcommand{\rmenys}{\rho_s^{(-)}}
\newcommand{\rmenyst}{\rho_s^{(-) T_b}}
\newcommand{\roq}{\rho_q}
\newcommand{\roqt}{\rho_q^{T_b}}
\newcommand{\rostb}{\rho^{T_b}}

\def\un{\leavevmode\hbox{\normalsize1\kern-4.6pt\large1}}

\begin{document}
\draft
\title {Quantum inseparability as local pseudomixture}
\author{Anna Sanpera$^{\dagger}$, Rolf Tarrach$^{\ddagger}$ and Guifr\'e Vidal$^{\ddagger}$}

\address{$^{\dagger}$CEA/DSM/DRECAM, Service des Photons, Atomes et Molecules, Centre d'Etudes de Saclay,   
\mbox{91191 Gif-Sur-Yvette,France.}}
\address{$^{\ddagger}$ Departament d'Estructura i Constituents de la Mat\`eria,
Universitat de Barcelona, 
\mbox{08028 Barcelona, Spain.}}

\date{\today}

\maketitle

\begin{abstract}
We show how to decompose any density matrix of the simplest binary composite systems, whether separable or not, in terms of only product vectors. We determine for all cases the minimal number of product vectors needed for such a decomposition. Separable states correspond to mixing from one to four pure product states. Inseparable states can be described as \em pseudomixtures \rm of four or five pure product states, and can be made separable by mixing them with one or two pure product states.

\end{abstract}

\vspace{0mm}

Entanglement, inseparability and nonlocality are some of the most genuine
quantum concepts. 
While 
for pure states it is well established since long ago that the non-local character of the composite system is revealed in different but equivalent ways, the situation is drastically
different for mixed states.
For example, for pure states the violation of some kind of Bell inequalities\cite{bell}, or the demonstration that no local hidden variable models can account for the correlations between the observables in each subsystem, are equivalent definitions of non-locality \cite{gisin}. But for mixed states, described by density matrices, such equivalences fade away. Consider a composite quantum system described by a density
matrix $\rho$ in the Hilbert space ${\cal H}_a\otimes {\cal H}_b$. 
In the frame set by the concepts of our
starting sentence, product or factorizable states are the simplest possible.
They are of the form $\rho_p=\rho_a\otimes\rho_b$, i.e. for them, and only for them, the description of the two isolated subsystems is equivalent to the description of the 
composite system. Recalling that subsystems are described by the reduced
density matrices obtained via partial tracing: $\rho_a=Tr_b\rho$ (
$\rho_b=Tr_a\rho$), a density matrix corresponds to a product or factorizable state 
iff 
\begin{equation}
\rho=Tr_b\rho\otimes Tr_a\rho  \iff \rho=\rho_p.
\end{equation}
Also their index of correlation (or mutual information) defined in terms of von Neumann entropies of the system and subsystems,
\begin{equation}
I_c=Tr\rho \ln\rho-Tr \rho_a \ln\rho_a -Tr \rho_b\ln\rho_b,
\end{equation}
vanishes, and this happens only for them\cite{bar}. Their subsystems are uncorrelated. Any state which is not a product state presents some kind
of correlation. They are called correlated states. 
Quantum mechanics has taught us that there is a hierarchy of 
correlations, and the physics in the different ranks is different.
The simplest correlated states are the classically correlated ones. Separable states are either uncorrelated or classically correlated. Their density matrices can always be written in the form:
\begin{equation}
\rho_s=\sum_{i} p_i \rho_{ai}\otimes\rho_{bi};\;\; 1\ge p_i > 0;\,\, \sum_ip_i=1,
\label{separable}
\end{equation}
i.e. as a mixture of product states. 
Their characterization is notoriously
difficult. Thus, given a density matrix which is known to describe
a separable state only very recently algorithms for decomposing it according to Eq. (\ref{separable})
have been found \cite{san,woot}; besides, the decomposition is not unique.
In fact, only recently Peres and the Horodecki family\cite
{per,ho3} have obtained a mathematical characterization of these states, at least when the dimension of the composite Hilbert space is $2\times 2$ or
$2\times 3$. For these cases the necessary and sufficient condition for separability is that the matrix obtained by partially transposing 
the density matrix $\rho$ is still a density matrix, i.e. with only non-negative eigenvalues
\begin{equation}
\rho^{T_b}=(\rho^{T_a})^*\ge 0 \iff \rho=\rho_s.
\label{per}
\end{equation}
For composite systems described by Hilbert spaces of higher dimensions, the positivity condition of $\rho^{T_b}$ is only a necessary one for separability\cite{ho3}. 
Following the hierarchy of correlations, we find the states that are no longer separable, i.e.  $\rho\not=\rho_s$. These states are called  ``EPR-states''\cite{epr}, ``inseparable'', ``non-local'', and sometimes ``entangled'' or simply ``quantum-correlated'' to emphasize that  their correlations are not
strictly classical anymore, though often these labels do not refer to exactly the same states. This confusion reflects the need of a further subclassification of the inseparable states according to whether they admit local hidden variables, whether they violate some kind of Bell inequality\cite{wer,ho21}, etc..

The issue we want to address here is whether any state, even if non-local, allows for some kind of local description. We will see that this leads to novel physical perspectives about non-locality. Thus the aim of this contribution is to decompose any separable or inseparable density matrix of a binary composite system of dimension 2 $\times$ 2 in terms of only product vectors, and to give for all cases the minimal number of product vectors needed. In other words, we give the minimal local description of any state, be it separable or not. (Here and in what follows "local" means that it refers to the subsystems). More specifically, we will start proving that any separable density matrix can always be written 
as: 
\begin{equation}
\ros=\sum_{i=1}^{n} p_i (\ket{e_i}\bra{e_i}\otimes\ket{f_i}\bra{f_i});\,1\geq p_i > 0;\, \sum_{i=1}^{n} p_i = 1,
\label{decpro}
\end{equation}
with $1 \leq n \leq 4$, and we will determine the minimal $n$ as a function of $\rho_s$. This introductory result completes the result $n \leq 5$ of \cite{san} and reproduces the result $n \leq 4$ of \cite{woot} in a completely independent way. Calling statistical mixtures of pure product states $\ket{e_i}\otimes\ket{f_i}$,  $\ket{e_i}\in{\cal H}_a$, $\ket{f_i}\in{\cal H}_b$, {\em local mixtures}, and calling the smallest $n$ its {\em cardinality}, Eq. (\ref{decpro}) says that any separable density matrix is a local mixture of cardinality smaller than five. We then come to our main results. First, any pure inseparable state  ($\roq=\roq^2$) can be written as:
\begin{eqnarray}
\roq &= &(1+q_1+q_2)\rmes \nonumber\\
&-& \sum_{i=1}^{2} q_i(\ket{g_i}\bra{g_i}\otimes\ket{h_i}\bra{h_i});\, 0<q_i <\infty,
\label{roqpur}
\end{eqnarray}
with $\rmes$ separable of cardinality 3. The subscript q means inseparable or quantum correlated. And, second, any non-pure inseparable state ($\roq>\roq^2$) can be written as:
\begin{equation}
\roq = (1+q)\rmes - q(\ket{g}\bra{g}\otimes\ket{h}\bra{h}); \, 0< q <\infty,
\label{roqmix}
\end{equation}
with $\rmes$ separable of cardinality 3 or 4. We finally determine the cardinality of $\rmes$ as a function of $\roq$.
As a consequence of our results any inseparable density matrix can be written as what we call a {\em pseudomixture}:
\begin{equation}
\roq = (1+q)\rmes - q\rmenys;\, 0<q<\infty,
\label{pseudomix}
\end{equation}
of cardinality $n \equiv n^{(+)} + n^{(-)}$,  $n^{(+)}$ and  $n^{(-)}$ being the cardinalities of $\rmes$ and $\rmenys$. Then, in a nutshell, our main result is to determine for any state its representation in form of a local (pseudo)mixture of minimal $n^{(-)}$ and then minimal $n^{(+)}$. Local pseudomixtures have an interesting physical interpretation. Eq.(\ref{roqmix}) for instance, says that any inseparable mixed state can be made separable by mixing it with some pure product state or, that its quantum correlations can be completely washed out with only one single local mixing preparation.

Before proving all this, let us mention that local pseudomixtures lead immediately to an unambiguous measure of entanglement,
\begin{equation}
E(\roq) = min\; q,
\end{equation}
where $q$ is defined in Eq.(\ref{pseudomix}). It is unambiguous because in Eq. (\ref{pseudomix}) only product states appear and thus $E(\roq)$ just represents the minimal local mixing needed to wash out all entanglement. Minimizing $q$ is however different from minimizing $n^{(-)}$ and then $n^{(+)}$, which is what we do here, and we postpone its study and comparison with other entanglement measures \cite{woot,bennett,vedr,lewen} for the time being.

In order to prove Eq.(\ref{decpro}) we need the following theorems:

\noindent{\bf Theorem1.} For any plane ${\cal P}_1$ in ${\cal C}^2\otimes {\cal C}^2$  defined by two product vectors $\ket{v_1}$ and $\ket{v_2}$,
either all the states in this plane are product vectors, or there is no other product vector in it.\\
\noindent Proof: With the help of $SU(2)\otimes SU(2)$ transformations, $\ket{v_1}$ and $\ket{v_2}$ can always be expressed so that:
\begin{equation}
{\cal P}_1\equiv\alpha_1{1\choose 0}\otimes{1\choose 0} + \beta_1 {\cos A \choose \sin A}\otimes {\cos B \choose \sin B},
\label{teor1}
\end{equation} 
with $0\leq A,B\leq \pi/2$; $A$ and $B$ not simultaneously vanishing and $\alpha_1, \beta_1 \in {\cal{C}}$. All vectors in ${\cal P}_1$ are  product vector iff
$\sin A\sin B=0$. If $\sin A\sin B\neq 0$, 
then the only product vectors contained in ${\cal P}_1$ are the generators of the plane
$\ket{v_1}$ and $\ket{v_2}$ . 

\noindent{\bf Corollary.} If $\rho$ has rank 2 and is separable it can
always be expressed as a statistical mixture of two pure product states and
thus $\rho^{T_b}$ is also of rank 2.\\
\noindent It suffices to see that for any separable $\rho$ of rank 2 its range,  ${\cal R}(\rho)$, is a plane of type ${\cal P}_1$. If it only contains two product vectors, then necessarily $\rho=p\ket{v_1}\bra{v_1}+(1-p)\ket{v_2}\bra{v_2}$ for some $0<p<1$. In the case that all vectors
in  ${\cal R}(\rho)$ are product vectors then its spectral decomposition gives us immediately the desired decomposition. Since in any case:
\begin{equation}
\rho=p\ket{e_1f_1}\bra{e_1f_1}+(1-p)\ket{e_2f_2}\bra{e_2f_2},
\end{equation}
it immediately follows that $\rostb$ is also of rank 2.

\noindent{\bf Theorem 2.} Any plane ${\cal P}_2$ in ${\cal C}^2\otimes {\cal C}^2$  contains at least one product vector. Some planes contain only one. \\
\noindent Proof: Consider the plane ${\cal P}_2$ generated by two orthogonal
vectors. Again, with the help of $SU(2)\otimes SU(2)$ transformations, it can
be expressed as 
\begin{equation}
{\cal P}_2\equiv \alpha_2
             \pmatrix{  A \cr
                        0     \cr
                        0      \cr
                        B   \cr} 
+ \beta_2 \pmatrix{C B \cr
                    \gamma \cr
                    \delta \cr
                  -C A\cr},
\label{teor2}
\end{equation}
with $A,B,C \in {\cal R}$ and $\gamma, \delta, \alpha_2, \beta_2\in {\cal C}$. Assume that none of the generating vectors is a product vector, that is $AB\neq 0$  
 and $C^2 AB+\gamma\delta\neq 0$. Then a vector in ${\cal P}_2$ is a product vector iff
\begin{equation}
\alpha_2^2 AB + \alpha_2\beta_2 C (B^2-A^2)-\beta_2^2(C^2 AB+\gamma\delta)=0.
\label{segongrau}
\end{equation}
With the above restrictions on $A,B,C,\gamma$ and $\delta$, there is always at least one non-vanishing solution (i.e. $\alpha_2, \beta_2$ such that $\alpha_2\beta_2\neq0$)  of Eq.(\ref{segongrau}), There is sometimes only one non-vanishing solution (see also \cite{grass}).

We can now outline our procedure for finding the decomposition of a separable state into four pure product states. We will first prove that five pure product states always do, and then present the slightly more cumbersome proof of going from five to four pure product states. The algorithm consists in subtracting a projector onto a product vector from $\rho_s$ or $\rho_s^{T_b}$ in such a way that $r(\rho_s)+ r(\rho_s^{T_b})$  diminishes at least in one unity (here $r(\rho)$ means the rank of $\rho$). We then repeat the procedure till the desired decomposition is obtained. Consider the most general case, a separable state $\ros$ such that both itself and its partially
transposed matrix are of rank 4: $r(\rho_s)=r(\rho_s^{T_b})=4$. As we shall see all the other cases are subcases of this one.
Define now
\begin{equation}
\rho(p)\equiv \frac{1}{1-p}(\rho_s-p\ket{e_1,f_1}\bra{e_1,f_1}); \,\,\,
0 < p < 1,
\label{p1}
\end{equation}
and
\begin{equation}
\rho(p)^{T_b} = \frac{1}{1-p}(\rho_s^{T_b}-p\ket{e_1,f_1^*}\bra{e_1,f_1^*});\,\,\,
0 < p < 1,
\label{p1t}
\end{equation}
where $\ket{e_1}\in{\cal H}_a$ and $\ket{f_1}\in{\cal H}_b$  are completely arbitrary states. For $p$ small enough both $\rho$ and $\rho^{T_b}$,
are positive, and therefore, due to Eq.(\ref{per}), separable. Let us
denote by $p_1$ the smallest value for which a zero eigenvalue appears in $\rho(p)$
or $\rho(p)^{T_b}$. Let us assume that for $p_1$ one eigenvalue of $\rho(p)$ is equal to zero, i.e. $r(\rho(p_1))=3$ and 
$r(\rho(p_1)^{T_b})=4$ (the same argument holds for the opposite case).
Consider now a new product vector belonging to the
range of $\rho(p_1)$, $\ket{e_2,f_2}\in{\cal R}(\rho(p_1))$,
and define a new density matrix:
\begin{equation}
\bar\rho(p)\equiv \frac{1}{1-p}(\rho(p_1)-p\ket{e_2,f_2}\bra{e_2,f_2});\,\,\, 0 < p < 1.
\label{p2}
\end{equation}
\noindent As before, for $p$ small enough, both $\bar\rho(p)$ and $\bar\rho(p)^{T_b}$ are non-negative and
thus separable. Let us denote by $p_2$ the smallest value of $p$ for which either
$\bar\rho(p)$  or $\bar\rho(p)^{T_b}$ develop a new vanishing eigenvalue. It cannot be $ \bar\rho(p) $ unless, because of the corollary, $ \bar\rho(p)^{T_b}$  develops simultaneously two vanishing eigenvalues. Therefore, it is in general $\bar\rho(p)^{T_b}$ which will develop a new vanishing eigenvalue, so that 
\begin{equation}
r(\bar\rho(p_2))=r(\bar\rho(p_2)^{T_b})=3.
\label{ran3}
\end{equation}
As $\bar\rho(p_2)$ has a decomposition of the type of Eq.(\ref{decpro}) 
with at least three terms, and $\bar\rho(p_2)^{T_b}$ has the corresponding partially transposed one,
there always exists a product state satisfying:
$\ket{e_3,f_3}\in{\cal R}(\bar\rho(p_2))$  and  $\ket{e_3,f_3^*}\in{\cal R}(\bar\rho(p_2)^{T_b})$\cite{hug,pawel}. 
Define now:

\begin{equation}
\tilde\rho(p)\equiv \frac{1}{1-p}(\bar\rho(p_2)-p\ket{e_3,f_3}\bra{e_3,f_3});\,\,\, 
0 < p < 1.
\label{p3}
\end{equation}
\noindent It is clear from the corollary  that a $p_3$ exists such that:
\begin{equation}
r(\tilde\rho(p_3)\geq 0)=r(\tilde\rho(p_3)^{T_b}\geq 0)=2,
\label{ran2}
\end{equation}
and then it immediately follows that:

\begin{eqnarray}
\tilde\rho(p_3)&\equiv& p_4\ket{e_4,f_4}\bra{e_4,f_4}\nonumber\\
           &+& (1-p_4)\ket{e_5,f_5}\bra{e_5,f_5}; \,\,\, 0 < p_4 < 1,
\label{p45}
\end{eqnarray}
completing thus the decomposition of any separable state. Therefore :
\begin{eqnarray}
\ros&=&p_1P_1+p_2(1-p_1)P_2\nonumber \\
&+&p_3(1-p_2)(1-p_1)P_3\nonumber \\
&+&p_4(1-p_3)(1-p_2)(1-p_1)P_4\nonumber \\
&+&(1-p_4)(1-p_3)(1-p_2)(1-p_1)P_5,
\label{p12345}
\end{eqnarray}
where $P_i\equiv\ket{e_i,f_i}\bra{e_i,f_i}$ are projectors onto pure product vectors. This proves Eq. (\ref{decpro}) with $n \leq 5$. Notice that if $r(\rho_s)+ r(\rho_s^{T_b}) < 8$ then $n < 5$.

Let us now show that even when  $r(\rho_s)+ r(\rho_s^{T_b}) = 8$ one can always find a decomposition into four pure product states instead of five. To do this, we shall prove that there exists always at least one projector $P=\ket{e,f}\bra{e,f}$  and its partially transposed $P^{T_b}=\ket{e,f^*}\bra{e,f^*}$ that can
be subtracted from $\rho_s$ and $\rho_s^{T_b}$  respectively in such a way 
that positivity is preserved and the rank of both matrices diminishes simultaneously by one unit. Let us proceed defining as in Eq. (\ref{p1}), but for each of the five product projectors of Eq. (\ref{p12345}), the following five matrices:
\begin{equation}
\rho_i(p)\equiv \frac{1}{1-p}(\rho_s-p\ket{e_i,f_i}\bra{e_i,f_i}); \,\,\,
0 < p < 1; \,i=1,...,5.
\end{equation}
We will fix two sets of five values of $p$ by the ten conditions

\begin{eqnarray}
r(\rho_i(p=s_i) \geq 0) = 3 \nonumber \\
r(\rho_i^{T_b} (p=\bar{s_i})\geq 0) = 3.
\end{eqnarray}
These conditions determine the maximal weights consistent with positivity with which 
the projectors  $P_i = \ket{e_i,f_i}\bra{e_i,f_i}$ and  $P_i^{T_b} = \ket{e_i,f^*_i}\bra{e_i,f^*_i}$ can be subtracted from $\rho_s$ and $\rho_s^{T_b}$ respectively.
We now show that it is impossible that $s_i < \bar{s_i}$ $\forall i$ or that $s_i > \bar{s_i}$ $\forall i$. From \cite{lewen} one knows the expressions for $s_i$ and $\bar{s_i}$ as defined above,
\begin{eqnarray}
s_i = \frac{1}{ \bra{e_i,f_i}\rho_s^{-1}\ket{e_i,f_i} }\nonumber \\
\bar{s_i} = \frac{1}{ \bra{e_i,f_i^*}(\rho_s^{T_b})^{-1}\ket{e_i,f_i^*}}
\label{sis}
\end{eqnarray}
If we call $p_i$ the probabilities for which $P_i$ appears in $\ros$ (cf. Eq. (\ref{decpro})), then if, say, $s_i < \bar{s_i}$ $\forall i$ it immediately follows
\begin{equation}
\sum_{i=1}^{5} p_i s_i^{-1} > \sum_{i=1}^{5} p_i \bar{s_i}^{-1},
\end{equation}
which from Eq. (\ref{sis}) reads:
\begin{equation}
\sum_{i=1}^{5} p_i \bra{e_i,f_i}\rho_s^{-1}\ket{e_i,f_i} > \sum_{i=1}^{5} p_i \bra{e_i,f_i^*}(\rho_s^{T_b})^{-1}\ket{e_i,f_i^*} ,
\end{equation}
or equivalently:
\begin{equation}
Tr(\rho_s\rho_s^{-1}) > Tr(\rho_s^{T_b}(\rho_s^{T_b})^{-1}),
\end{equation}
which cannot be. Thus at least for one $i$, say $j$, $s_j \geq \bar{s_j}$. If they are equal, then subtracting this $\ket{e_j,f_j}\bra{e_j,f_j}$ from $\ros$ in Eq. (\ref{p1}) allows to reach 
\begin{equation}
r(\rho(s_j) \geq 0) = r(\rho(s_j)^{T_b} \geq 0)=3
\label{enquatre}
\end{equation}
in one step. If  $s_j > \bar{s_j}$ then by connectivity of the space of product vectors and continuity of $s$ and $\bar{s}$ as defined by Eq. (\ref{sis}) as functions of the states of this space, there exists one  $\ket{e,f}\bra{e,f}$ which has $s = \bar{s}$ and for which Eq. (\ref{enquatre}) holds. Thus always a decomposition with four terms exists, and Eq. (\ref{decpro}) has been proven with 
\begin{equation}
n = max( r(\ros),  r(\rho_s^{T_b}) ) \leq 4.
\label{nderho}
\end{equation}

Let us now obtain our main results, which refer to inseparable states. From Eq. (\ref{per}) we know that 
\begin{equation}
{\it inf}\, \sigma(\rho^{T_b})<0 \Longleftrightarrow \rho=\rho_q,
\end{equation}
where $\sigma(\rho)$ means the spectrum of $\rho$. Let us prove that $\rho^{T_b}$ has only one negative eigenvalue. If there were two one could always find, according to theorem 2, a product vector  $\ket{e,f}$ in the plane defined by the corresponding two eigenvectors, and for which obviously
\begin{equation}
\bra{e,f}\rho_q^{T_b}\ket{e,f}<0.
\end{equation}
But the above expression is equivalent to
\begin{equation}
\bra{e,f^*}\rho_q\ket{e,f^*}<0,
\end{equation}
which is impossible, since $\roq \geq 0$. We will call the eigenvector of negative eigenvalue $\ket{N}$, i.e.,
\begin{equation}
\roq^{T_b}\ket{N} = -N\ket{N};\, N > 0.
\end{equation}
We will now see that $\roq$ can be made separable by mixing it statistically with an adequate separable density matrix, $\rmenys$, i.e.:
\begin{equation}
\rho(q) \equiv \frac{1}{1+q} (\roq + q \rmenys),
\end{equation}
where $ 0<q<\infty$ is such that
\begin{equation}
\rho(q)^{T_b} = \frac{1}{1+q} (\roqt + q \rmenyst) \geq 0. 
\label{insep}
\end{equation}
We want to do this in a doubly minimal way. We want to choose $\rmenys$ to have minimal rank, and we then want to choose the minimal $q$, i.e. such that $\rho(q)^{T_b}$ just develops a vanishing eigenvalue ($r(\rho(q)^{T_b})<4$). Notice that due to the Hellmann-Feynman theorem \cite{fey} the only eigenvalue of $\roqt$ which can become zero by adding a non-negative operator is its negative eigenvalue.

\noindent We will show how this is done as a function of the rank of $\roq$:

1. Assume $r(\roq) = 1$.
Here $\roq$ represents an entangled pure state, which can always be written with the help of the $SU(2)\otimes SU(2)$ transformations in its canonical form (cf. Eq. (\ref{teor2})) $\bra{\epsilon} \equiv (\cos A, 0, 0, \sin A)$ with $\cos A \sin A > 0$. It turns out that $\bra{N} = \frac{1}{\sqrt{2}} (0, 1, -1, 0)$, and that $r(\roqt) = 4$, as $\sigma(\roq^{T_b})$ = $\{ \cos^2 A$, $ \sin^2 A$, $ \cos A\sin A$, $ -\cos A\sin A (= - N)\}$. So, in this case, the minimal $q$ satisfies $r(\rho(q)^{T_b})=3$. This implies that the rank of $\rmenys$ cannot be one. Indeed, if it were one, as $r(\roq)=1$ it would imply $r(\rho(q))=2$. But the two conditions $r(\rho(q)^{T_b})=3$ and $r(\rho(q))=2$ cannot be simultaneously satisfied for a separable density matrix (cf. corollary). On the other hand a $\rmenys$ with $r(\rmenys) = 2$ which does the job can always be found. It leads to $r(\rho(q)^{T_b}) = r(\rho(q)) = 3$. It can be implemented by choosing the two product vectors which statistically mixed represent $\rmenys$ to be the vectors $\ket{g_i,h_i}$ given by the Schmidt decomposition of $\ket{N}$, $\ket{N} = c_1 \ket{g_1,h_1^*} + c_2 \ket{g_2,h_2^*}$. This proves Eq. (\ref{roqpur}) with $\rmes = \rho(q),\; q = q_1 + q_2$ and where the result of Eq. (\ref{nderho}) shows that the cardinality of is $\rmes$ is 3.

2. Assume $r(\roq) = 2$.
Taking $\ket{e,f} \in {\cal R}(\roq)$ which by theorem 2 always exists, we write $\roq$ in the form \cite{lewen}
\begin{equation}
\roq = \frac{1}{1+p}(\ket{\Psi}\bra{\Psi}+p\ket{e,f}\bra{e,f}); \,\,\, p > 0,
\label{roqrank2}
\end{equation}
where $\ket{\Psi}$ is an entangled vector which belongs to ${\cal R}(\roq)$. Let us now prove that $r(\roqt) = 4$. In order to do so write $\ket{\Psi}$ in its canonical form $\ket{\epsilon}$. Consider the partially transposed of Eq. (\ref{roqrank2}). Recall (from the previous case) that $\ket{\epsilon}\bra{\epsilon}^{T_b}$ has three positive and one negative eigenvalues. The negative eigenvalue cannot be made to vanish adding the non-negative operator $\ket{e,f^*}\bra{e,f^*}$ because then $\roqt \geq 0$, which from Eq. (\ref{per}) is inconsistent with $\roq$ being inseparable. This, recalling that positive eigenvalues certainly cannot be made to vanish, proves $r(\roqt) = 4$.
This, in fact, always holds, so that $r(\roqt) = 4$ independently of $r(\roq)$. 
It is now not too difficult to show that for any $\ket{e,f}$ always at least one $\rmenys \equiv \ket{g,h}\bra{g,h}$ exists which allows to satisfy Eq. (\ref{insep}) with $r(\rho(q)) = r(\rho(q)^{T_b}) = 3$. The upshot of this is that Eq. (\ref{roqmix}) holds with $\rmes = \rho(q)$ of cardinality 3.

3. Assume $r(\roq) = 3$.
As the previous case always allowed to find a $\rmenys$ with $r(\rmenys) = 1$ this is {\em a fortiori} true now too. This proves Eq. (\ref{roqmix}), but it is now not obvious whether it can always be done with a $\rmes$ of cardinality 3. In fact, it cannot, as the analysis of the following counterexample shows:
\begin{equation}
\roq = \frac{1}{1+p_1+p_2}(\ket{\Psi}\bra{\Psi}+p_1\ket{e_1,f_1}\bra{e_1,f_1} + p_2\ket{e_2,f_2}\bra{e_2,f_2}); \,\,\, p_i > 0,
\end{equation}
with $\ket{\Psi} =\ket{\epsilon}$, $\bra{e_1} = \bra{f_i} = (1,0)$ and $\bra{e_2} = (0,1)$. Indeed, none of the $\ket{g,h}$ vectors belonging to ${\cal R}(\roq)$, which either have $\ket{g} = \ket{e_2}$ or $\ket{h} = \ket{f_i}$, does the job, and thus $r(\rho(q)) = 4$. On the other hand it is easy to find examples of $\roq$ for which $r(\rho(q)) = r(\rho(q)^{T_b}) = 3$. Thus Eq. (\ref{roqmix}) is proven but $\rmes$ does not have always cardinality 3. This parallels the ambiguity of $n$ for separable states of rank 3, for which also sometimes $n=3$ and sometimes $n=4$.

4. Finally assume $r(\roq) = 4$. In this case, obviously Eq. (\ref{roqmix}) holds for $\rmes$ of cardinality 4.

\vspace*{3mm}

To summarize, we have proven that any separable state in ${\cal C}^2\otimes {\cal C}^2$ is a local mixture of at most cardinality four, that any inseparable state in ${\cal C}^2\otimes {\cal C}^2$ is a local pseudomixture of cardinality four or five and that any inseparable state can be made separable by mixing it with only one single pure product state, except if it is pure, in which case it needs to be mixed with two pure product states. Therefore, when a state has only quantum correlations these can be made classical by mixing it with two pure product states, while when it has both classical and quantum correlations, mixing it with one single pure product state suffices to wash out all quantum correlations.

\vspace*{3mm}

We are specially grateful to M. Lewenstein for helping us to prove the decomposition
of a separable state into just four product vectors (Eq. 24-28). A.S also thanks P. Horodecki and A. Peres for useful discussions and acknowledges financial support from EC. R.T. enjoys financial support by CICYT (Spain), grant AEN95-0590 and by CIRIT (Catalonia), grant 1996SGR-00066. G.V. acknowledges a CIRIT grant 1997FI-00068 PG.

\end{document}